\documentstyle[aps,multicol,prb]{revtex}
\renewcommand{\narrowtext}{\begin{multicols}{2} \global\columnwidth20.5pc}
   \renewcommand{\widetext}{\end{multicols} \global\columnwidth42.5pc}
  \newcommand{\wide}{\widetext \noindent \line(200,0){245} \line(0,1){3}\\}
  \newcommand{\narrow}{\begin{flushright}\mbox{\line(0,-1){3}$\! \!$
  \line(1,0){245}} \end{flushright} \narrowtext \noindent}
\begin{document}
\draft
\title{Nonequilibrium orbital magnetization of strongly localized electrons}
\author{Y. Galperin$^{(a)}$ and O. Entin-Wohlman$^{(b)}$}
\address{$^{(a)}$Department of Physics, University of Oslo, P. O. Box 1048
Blindern, 0316 Oslo, Norway \\and A. F. Physico-Technical Institute,
194021 St. Petersburg, Russia \\ 
$^{(b)}$School of Physics and Astronomy, Raymond and Beverly Sackler
Faculty of Exact Sciences, \\
Tel Aviv University, Tel Aviv 69978, Israel}

\date{\today}
\maketitle

\begin{abstract}
The magnetic response of strongly localized electrons to a
time-dependent vector 
potential is considered. The orbital magnetic moment 
of the system, away from steady-state conditions,  
is obtained. The expression involves the tunneling and phonon-assisted 
hopping currents between localized
states. The frequency and 
temperature dependence of the orbital magnetization
is analyzed as function of the admittances connecting localized levels. 
It is shown that quantum interference
of the localized wave functions contributes to the moment a term which follows
adiabatically the time-dependent perturbation.
\end{abstract}

\narrowtext

\section{Introduction}

The explanation of the orbital magnetic moment at thermal equilibrium
in terms of ``ring 
currents" has been known  
for more than half a century \cite{pauling}. 
The most studied configuration is that of a small mesoscopic 
structure penetrated by a constant magnetic flux, 
in which the electrons move ballistically or diffusively \cite{buttiker}.
Recently, the problem of the ring currents and the associated magnetic moment
have been addressed in the case of strong disorder \cite{we}.
In this regime the electronic states are
localized and their overlaps are small. It has been shown that since
the coupling of the electrons to  
the thermal bath enables resonant processes 
\cite{holstein}, there appears a current flowing in opposite 
sense to the one due to tunneling, which exists in the 
absence of the electron-phonon interaction. This counter
current has been related  
to the dc Hall effect in insulators \cite{we2}. 

The study of  the orbital magnetic response  to a {\it time-dependent}
flux \cite{kohn,landauer} is of interest by itself and also because it
is relevant to experiments \cite{levy}  designed to detect  persistent
currents   in  mesoscopic  rings   \cite{triveldi}.     Nonequilibrium
magnetization  may be used   to measure specific electronic relaxation
rates \cite{galperin}.     The  magnetic  moment   in  response  to  a
time-dependent flux  has been discussed so far  in the context of weak
disorder. 

In  this paper we  investigate    the response of strongly   localized
electrons to a time-dependent flux, which sets up a magnetic moment in
the system.  We obtain  the nonequilibrium orbital   magnetization and
study it when the electrons are coupled to the thermal bath. 

The  calculation  is  based on  the  Kubo  approach  that yields,  for
strongly localized electrons,   the  current between   two
localized  states   in  response  to  the   time-dependent   flux. The 
proper definition of the orbital magnetization away from steady-state 
conditions, in terms of these currents, is a somewhat  delicate
point. We devote  the next section to  a discussion of  this issue. In
Sec. III we obtain the  current that flows
between two localized states. Section IV contains the analysis of the
characteristic frequencies that dominate the nonequilibrium magnetization 
of an isolated group of three localized states. (This is the smallest
cluster which can have an orbital moment.) Section V contains  
concluding comments.

The  formalism we    develop allows for   the calculation   of the  ac
conductance of strongly localized electrons coupled to a thermal bath.
We  derive in the Appendix quantum interference corrections to the 
ac hopping conductance and show  that they yield nonlocal contributions
to the `bond' admittances. 

\section{Nonequilibrium magnetization of localized electrons}

Our aim is to obtain an expression for the orbital magnetic moment 
of localized electrons, at nonequilibrium conditions. We will show that 
the moment can be given in terms of  the
currents between the localized  states.  When  these currents
flow in    response  to a {\it constant}     magnetic field, i.e.   at
equilibrium,  this is  a   straightforward  task.  In  that  case  the
equilibrium value of the orbital magnetization, $\vec{M}_{\text{eq}}$,
is just  a manifestation   of  the  Aharonov-Bohm  effect and  can  be
obtained from the thermodynamic potential \cite{we}. The result reads 
\begin{equation}
\vec{M}_{\text{eq}}=-\frac{1}{2c}\sum_{\langle \ell j\rangle}I_{\ell
j}[\vec{R}_{\ell}\times\vec{R}_{j}],\label{M1} 
\end{equation}
in which $I_{\ell j}$ is  the current between  the state localized  at
$\vec{R}_{\ell}$ and  the   state  localized  at   $\vec{R}_{j}$,  and
$\langle \ell
j\rangle$ indicates a sum over pairs. At equilibrium, current conservation
implies   that  the     `bond'   currents    $I_{\ell   j}$    satisfy
$\sum_{\ell}I_{\ell j}=0$,      which    in    turn     yields    that
$\vec{M}_{\text{eq}}$ is independent of  the choice of the  coordinate
origin. At nonequilibrium one has 
\begin{equation}
\dot{\rho}_{j}=\sum_{\ell}I_{\ell j}, \label{M2}
\end{equation}
where $\dot{\rho}_{j}$ is the time derivative of the electron
occupation on the $j$th  
localized state. The contribution of the dielectric currents resulting
from the nonequilibrium  
occupations has to be included in the orbital magnetic moment. For an
homogeneous continuous 
system, this is achieved \cite{panovsky}
by introducing a single vector function that encompasses both the
(spatial-dependent) 
charge and the current densities. In the present case, the analogous
quantity is 
\begin{equation}
{\cal J}_{\ell j}=I_{\ell j}+\frac{1}{{\cal
N}}(\dot{\rho}_{\ell}-\dot{\rho}_{j}),\label{M3} 
\end{equation}
where ${\cal N}$ is the number of localized states. This is so because the
generalized currents ${\cal J}_{\ell j}$ satisfy
\begin{equation}
\sum_{\stackrel{\ell}{\ell \neq j}}{\cal J}_{\ell j}=0\label{M4}
\end{equation}
which follows from the requirement of charge conservation
$(\sum_{j}\dot{\rho}_{j}=0)$.  
Note that the
requirement (\ref{M4}) corresponds to a current density with zero
divergence in the case of a homogeneous continuous system. As a
result, the nonequilibrium 
orbital magnetic moment, $\vec{M}$, is given by
\begin{equation}
\vec{M}=-\frac{1}{2c}\sum_{<\ell j>}{\cal J}_{\ell
j}[\vec{R}_{\ell}\times\vec{R}_{j}]\label{M5} 
\end{equation} 
and is also independent of the choice of the coordinate origin. It
takes a particularly 
convenient form when the relative coordinates $\vec{r}_{\ell}$
\begin{equation}
\vec{r}_{\ell}=\vec{R}_{\ell}-\vec{R}, \ \ 
\vec{R}=\frac{1}{{\cal N}}\sum_{j}\vec{R}_{j},\label{M6}
\end{equation}
are introduced
($\vec{R}$ is the coordinate of the center of gravity). Then, using
Eqs. (\ref{M2}) and (\ref{M3}) we obtain 
\begin{equation}
\vec{M}=-\frac{1}{2c}\sum_{\langle \ell \rangle}I_{\ell j}[\vec{r}_{\ell
}\times\vec{r}_{j}].\label{M7} 
\end{equation}
In this expression the radius-vectors $\vec{R}_{\ell }$
appearing in (\ref{M1}) are replaced by the relative coordinates
$\vec{r}_{\ell}$. It is seen that for the choice (\ref{M6}) the displacement
currents disappear from the expression for $\vec{M}$.

\section{The linear response expression for the current}

The properties of strongly localized electrons interacting with a
phonon bath are 
most conveniently discussed using the Hamiltonian introduced by
Holstein \cite{holstein1}. 
In that description the electron site density is coupled to the 
ion displacements. (The term `site'
denotes a localized state.) The Holstein Hamiltonian, in the presence of 
a vector potential, can be written in the form
\begin{equation}
{\cal H}=\sum_{\ell}\epsilon_{\ell}c_{\ell}^{\dagger}c_{\ell}+\sum_{\vec{q}}
\omega_{\vec{q}} b_{\vec{q}}^{\dagger}b_{\vec{q}}
+\sum_{j\ell}J_{\ell j}Q_{\ell j}e^{i\phi_{\ell
j}}c_{\ell}^{\dagger}c_{j},\label{1} 
\end{equation}
in which $Q_{\ell j}$ results from the electron-phonon interaction
\begin{equation}
Q_{\ell j}=\exp\left[-\sum_{\vec{q}}\frac{v_{\vec{q}}^{\ell
j}}{\omega_{\vec{q}}} 
(b_{\vec{q}}-b_{-\vec{q}}^{\dagger})\right],\ \ 
v_{\vec{q}}^{\ell j}=v_{\vec{q}}^{\ell}-v_{\vec{q}}^{j}.\label{2}
\end{equation}
Here $c_{\ell}$ and $c_{\ell}^{\dagger}$ are the annihilation and
creation operators of an electron 
in the state localized on site $\ell $ and $\epsilon_{\ell}$ are the
on-site energies, assumed to 
be randomly distributed (energies are measured from the Fermi
level). The phonon operators are 
denoted $b_{\vec{q}}$ and $b_{\vec{q}}^{\dagger}$, and
$\omega_{\vec{q}}$ is the phonon energy.  
In Eq. (\ref{2}), $J_{\ell j}$ is the overlap of two electronic wave
functions localized on sites 
$j$ and $\ell $, and $v_{\vec{q}}^{\ell} $ is the matrix element for
the electron-phonon interaction  
on site $\ell $. The strong localization regime is characterized by
$|  J_{\ell 
j}| \ll|\epsilon_{\ell}-\epsilon_{j}|$. The transfer of an electron
between two sites,  
as described by the last term of ${\cal H}$, can be accomplished by
tunneling, or can be  
accompanied by the exchange of one or more phonons with the thermal
bath (as is evident by  
expanding $Q_{\ell j}$ in terms of $v_{\vec{q}}^{\ell j}$). 

In this description the vector potential applied to the system appears
as a phase factor, 
$\exp i\phi_{\ell j}$, in which $\phi_{\ell j}$ is the time-dependent
flux (in units of the 
flux quantum $\phi_{0}$) contained in the area $[\vec{R}_{\ell
}\times\vec{R}_{j}]/2$, where 
$\vec{R}_{\ell }$ is the radius vector to the site $\ell $. The effect
of this flux is treated as a  
perturbation. Accordingly, we re-write the Hamiltonian (\ref{1}) in the form
\begin{equation} 
{\cal H}={\cal H}_{\text e}+{\cal H}_{ph}+{\cal H}_{1}+{\cal H}_{\text
{per}}, \label{3} 
\end{equation}
in which
\begin{eqnarray}
{\cal H}_{\text
e}&=&\sum_{\ell}\epsilon_{\ell}c_{\ell}^{\dagger}c_{\ell}, \ \ \  
{\cal
H}_{ph}=\sum_{\vec{q}}\omega_{\vec{q}}b_{\vec{q}}^{\dagger}b_{\vec{q}},
\nonumber\\
{\cal H}_{1}&=&\sum_{\ell j}J_{\ell j}Q_{\ell
j}c_{\ell}^{\dagger}c_{j},\label{4} 
\end{eqnarray}
refer to the unperturbed system, and
\begin{equation}
{\cal H}_{\text {per}}=\sum_{\ell j}J_{\ell j}Q_{\ell j}i\phi_{\ell
j}(t)c_{\ell}^{\dagger}c_{j}.\label{5} 
\end{equation}

The current flowing in response to the
time-dependent flux can be obtained by using   
the expression \cite{we}
\begin{equation}
\hat{I}_{j\ell }=2e\Im J_{\ell j}Q_{\ell j}e^{i\phi_{\ell
j}}c_{\ell}^{\dagger}c_{j} \label{6} 
\end{equation}
for the current operator. Then linear response theory yields
that the net current between sites $j$ and $\ell $ is 
\begin{eqnarray}
I_{j\ell}(t)&=&\int_{-\infty}^{t}dt_{1}\left< \Bigl[{\cal H}_{\text
{per}}(t_{1}),2\Im J_{\ell j}Q_{\ell j} 
(t)c_{\ell}^{\dagger}(t)c_{j}(t)\Bigr]\right>
\nonumber\\
&&+\phi_{\ell j}(t)\langle 2\Re J_{\ell j}Q_{\ell
j}(t)c_{\ell}^{\dagger}(t)c_{j}(t)\rangle.\label{7} 
\end{eqnarray}
In this expression the time dependences of the operators, as well as
the thermal averages 
(indicated by the angular brackets) are determined by the Hamiltonian
${\cal H}_{\text e}+{\cal H}_{\text {ph}}+{\cal H}_{1}$, Eqs. (\ref{4}).

We now derive an explicit expression for the current, valid in the
strong localization regime. 
As this regime is characterized by the small parameter 
$|J_{j\ell}|/|\epsilon_{j}-\epsilon_{\ell}| $, we expand
$I_{j\ell}$ in powers  
of the wave-function overlaps. Clearly, [{\it cf.} Eqs. (\ref{5}) 
and (\ref{7})] the expansion starts at
order $J^{2}$. Quantum interference effects appear first
\cite{holstein} in order 
$J^{3}$, which allows 
for the interference of the direct amplitude 
$j\rightarrow\ell $ with the indirect one, via a
third site $k$. Such processes lead to the appearance
of ``ring currents"  
for a group of at least three
sites in the case of a constant (time-independent) flux \cite{we}. 
In this paper we concentrate
on effects resulting from the second order in the overlaps. However,
to demonstrate 
that quantum interference effects of strongly localized electrons can be 
studied using the Kubo approach, we present in the Appendix the
derivation of the 
bond current to third order in the overlaps. 

To second order in the wave-function overlaps the bond current is given by
\wide
\begin{eqnarray}
I_{j\ell}^{(2)}
&=&  e|J_{j\ell}|^{2}\int_{-\infty}^{t}dt_{1}\phi_{j\ell}(t_{1})
\Bigl(n_{j}(1-n_{\ell})
2\Im e^{i\epsilon_{\ell j}(t-t_{1})}\langle Q_{j\ell}(t_{1})Q_{\ell
j}(t)\rangle 
- n_{\ell}(1-n_{j})2\Im e^{i\epsilon_{\ell j}(t-t_{1})}\langle Q_{\ell
j}(t)Q_{j\ell}(t_{1}) \rangle\Bigr) 
\nonumber\\
&-&e| J_{j\ell}| ^{2}\phi_{\ell j}(t)\int_{-\infty}^{t}dt_{1}
\Bigl(n_{j}(1-n_{\ell})2 \Im e^{i
\epsilon_{\ell j}(t-t_{1})}\langle Q_{j\ell}(t_{1})Q_{\ell
j}(t)\rangle
-n_{\ell}(1-n_{j})2\Im e^{i\epsilon_{\ell j}(t-t_{1})}\langle Q_{\ell
j}(t)Q_{j\ell}(t_{1}) 
\rangle\Bigr),\label{8}
\end{eqnarray}
\narrow
where we have used $\phi_{j\ell}=-\phi_{\ell j}$. (The superscript (2)
indicates a result to order $J^{2}$.) 
Here $n_{j}$ denotes the equilibrium electron occupation of site $j$, 
$n_{j}=(e^{\beta\epsilon_{j}}+1)^{-1}$ and 
$\epsilon_{\ell j}=\epsilon_{\ell}-\epsilon_{j}$. 

The time-dependent correlators of the phonon
operators $Q_{j\ell}$ are calculated with respect to the free phonon
Hamiltonian, ${\cal H}_{\text {ph}}$. 
One finds
\begin{eqnarray}
& &\langle Q_{\ell j}(t)Q_{j\ell}(t_{1})\rangle
=e^{2g(t-t_{1})-2g(0)},\nonumber\\ 
& &g(t)=\sum_{\vec{q}}\frac{|
v_{\vec{q}}|^{2}}{\omega_{\vec{q}}^{2}}\Bigl[(1+N_{\vec{q}})
e^{-i\omega_{\vec{q}}t}+N_{\vec{q}}
e^{i\omega_{\vec{q}}t}\Bigr],\label{9}
\end{eqnarray}
where $N_{\vec{q}}=(e^{\beta\omega_{\vec{q}}}-1)^{-1}$ is the Bose
function. Here we have taken 
$v_{\vec{q}}^{j}v_{-\vec{q}}^{\ell}\sim\delta_{j\ell}|
v_{\vec{q}}|^{2}$, which is based  \cite{holstein1} on
$v_{\vec{q}}^{\ell}=v_{\vec{q}}\exp(i\vec{q}\cdot\vec{R}_{\ell})$. 
We now insert Eq. (\ref{9}) into Eq. (\ref{8}) and put
\begin{equation}
\phi_{j\ell}(t)=\bar{\phi}_{j\ell}e^{i\omega t},\label{10}
\end{equation}
where $\bar{\phi}_{j\ell}$ is the amplitude of the ac flux
$\phi_{j\ell}$. It then follows that 
\begin{equation}
I_{j\ell}^{(2)}(t)=\bar{I}_{j\ell}^{(2)}(\omega)e^{i\omega t},\label{11}
\end{equation}
with the amplitude $\bar{I}_{j\ell}^{(2)}$ given by
\begin{eqnarray} \label{cc1}
\bar{I}_{j\ell}^{(2)}(\omega)&=&e\bar{\phi}_{j\ell}|
J_{j\ell}|^{2}\int_{0}^{\infty}dt\Bigl(1-e^{-i\omega t}\Bigr)
\nonumber\\
& &\times \Bigl[n_{j}(1-n_{\ell})2\Im e^{i\epsilon_{\ell
j}t+2g(-t)-2g(0)}
\nonumber\\ 
&&\quad -n_{\ell}(1-n_{j})2\Im e^{i\epsilon_{\ell
j}t+2g(t)-2g(0)}\Bigr].\label{12} 
\end{eqnarray}
One notes that $\bar{I}_{j\ell}^{(2)}(\omega)$ vanishes as the
frequency tends to zero. In other words, 
to order $J^{2}$ the current is driven by the time dependence of the
flux. This is not  
the case in higher orders in the overlaps. We show in the Appendix
that due to 
quantum interference of the electron wave functions (which appear
first to order 
$J^{3}$), the bond currents and consequently the orbital magnetic
moment have a part  
that follows {\it adiabatically} the external flux.

One can now define an effective admittance $z_{j\ell}^{(2)}$ for the
$j\ell $ bond 
\begin{equation}
\bar{I}_{j\ell}^{(2)}(\omega)=i\omega(\bar{\phi}_{j\ell}/e)z_{j\ell}^{(2)}
(\omega).\label{13}
\end{equation}
For weak electron-phonon coupling, the real part of the admittance becomes
\begin{equation}
\Re z_{j\ell}^{(2)}=e^{2}| J_{jl}|^{2}n_{j}
(1-n_{\ell})\gamma_{j\ell}(\omega ),\label{14}
\end{equation}
with
\begin{eqnarray}
\gamma_{j\ell}(\omega )&=&\frac{1}{2\omega }\Bigl[(1-e^{-\beta\omega})
\int_{-\infty}^{\infty}dte^{i(\epsilon_{j\ell}+\omega )t}(1+g(t))\nonumber\\
&+&(e^{\beta\omega}-1)\int_{-\infty}^{\infty}dte^{i(\epsilon_{j\ell}-\omega
)t} 
(1+g(t))\Bigr],\nonumber\\
\gamma_{j\ell}(\omega )&=&e^{\beta\epsilon_{j\ell}}\gamma_{\ell
j}(\omega ).\label{15} 
\end{eqnarray}
(It is assumed that the localized levels are not accidentally
degenerate.) It is seen 
that two types of transitions between the levels $\ell $ and $j$
contribute to 
$\Re z_{j\ell}^{(2)}$, resonant transitions and non-resonant,
phonon-assisted ones 
that exist due to the coupling with the thermal bath. At low
frequencies such that 
$\beta\omega\ll 1$ we have
\begin{eqnarray}
\Re z_{j\ell}^{(2)}&=&e^{2}\beta |
J_{j\ell}|^{2}n_{j}(1-n_{\ell})\pi\Bigl[\delta 
(\epsilon_{j\ell}+\omega )+\delta (\epsilon_{j\ell}-\omega )\Bigr]\nonumber\\
&+&G_{j\ell},\label{16} 
\end{eqnarray}
with $G_{j\ell}$ being the dc bond conductance. At low temperatures 
such that $\beta |\epsilon |\gg  1$, it becomes \cite{ambegaokar}
\begin{equation}
G_{j\ell}=e^{2}\beta | J_{j\ell}|^{2}\frac{v^{2}{\cal D}(|\epsilon_{j\ell}|)}
{\epsilon_{j\ell}^{2}}e^{-\frac{\beta}{2}[|\epsilon_{j}|
+|\epsilon_{\ell}|+|\epsilon_{j\ell}| ]}.\label{17}
\end{equation}
Here $v$ is the interaction matrix element for the electron-phonon
coupling, and  
${\cal D}$ is the phonon density of states. The resonant transitions,
described by  
the first term in (\ref{16}) require that the probability to find a
resonating 
pair of sites will not vanish. In a bulk system this mechanism may
therefore be expected 
to contribute. The full quantum-mechanical analysis of an isolated
pair of sites, valid for arbitrary frequencies, is given e.g.
in Ref. ~\onlinecite{maleev}.

The imaginary part of $z_{j\ell}^{(2)}$
is associated with the dielectric response of the $j\ell $ bond, and
is out-of-phase 
with the real term. In the absence of the coupling to the thermal bath
it becomes 
\begin{equation}
\Im z_{j\ell}^{(2)}(\omega )=e^{2}| J_{j\ell}| ^{2}
\frac{n_{j}-n_{\ell}}{\epsilon_{\ell j}}
\frac{\omega}{\epsilon_{\ell j}^{2}-\omega^{2}}.\label{18}
\end{equation} 
The coupling with the phonons
introduces additional contributions to the dielectric
response which are ignored here for simplicity.

\section{Analysis of the magnetic moment}

In exploiting the Kubo formula, one should interpret the driving force as an
{\it effective} one, which arises from the externally applied field
{\it and} the fields induced 
in the system. This means that the time derivative of the flux
should incorporate the changes in the local chemical
potentials \cite{ambegaokar}. 
Using the relative coordinates introduced in (\ref{M6}) one thus has
\begin{eqnarray}
i\omega\bar{\phi}_{j\ell}&=&i\omega\frac{e}{2c}\vec{H}
\cdot[\vec{r}_{j}\times\vec{r}_{\ell}]
\nonumber\\
&&- i\omega\frac{e}{2c}{[\vec{H}\times\vec{R}]}\cdot
(\vec{r}_{j}-\vec{r}_{\ell}) 
+\delta\mu_{j}-\delta\mu_{\ell},\label{C1}
\end{eqnarray}
in which $\delta\mu_{j}$ is the change of the chemical potential at site $j$.
The first term here results from the time dependence of the magnetic
field $\vec{H}$. 
In fact, $(e/2c)\vec{H}\cdot[\vec{r}_{j}\times\vec{r}_{\ell}]$ is just the
magnetic flux (divided by the flux quantum $\phi_{0}$) acquired from the
magnetic field between sites $j$ and $\ell $. The
second term can be interpreted as arising from an electric field,
 $\vec{E}=i\omega[\vec{H}\times\vec{R}]/2c$.
Accordingly, the last three terms in (\ref{C1}) represent the local 
electro-chemical potential difference.

To determine the local chemical potentials, and consequently the bond
currents, we need to 
solve the set of equations (\ref{M2}), where
\begin{equation}
\dot{\rho}_{j}=i\omega e \beta n_{j}(1-n_{j})\delta\mu_{j},\label{C2}
\end{equation}
and $I_{\ell j}$ is given by Eq. (\ref{13}). (We omit the superscript
$(2)$ for brevity.) 
Thus we have
\begin{eqnarray}
i\omega C_{j}\delta\mu_{j}&=&\sum_{\ell }z_{\ell j}\Bigl (i\omega\frac{e}{2c}
\vec{H}\cdot[\vec{r}_{\ell}\times\vec{r}_{j}]
\nonumber\\
&-&i\omega\frac{e}{2c}[\vec{H}\times\vec{R}]\cdot (\vec{r}_{\ell}-\vec{r}_{j})
+\delta\mu_{\ell}-\delta\mu_{j}\Bigr ),\label{C3}
\end{eqnarray}
where
\begin{equation}
C_{j}=e^{2}\beta n_{j}(1-n_{j})\label{C4}
\end{equation}
is the capacitance connected to the $j$th site. As the calculation is
restricted to 
terms linear in the magnetic field $\vec{H}$ the second term on the
right hand side 
of (\ref{C3}) can be omitted since it vanishes upon averaging over the
site locations. 
It follows that the local chemical potentials and consequently the
currents are 
determined by the magnetic fluxes $\varphi_{\ell j}$
\begin{equation}
\varphi_{\ell
j}=\frac{e}{2c}\vec{H}\cdot[\vec{r}_{\ell}\times\vec{r}_{j}].\label{C5} 
\end{equation}

The magnetic moment, Eq. (\ref{M7}), is given in terms of the bond
currents. To find those, 
we use Eqs. (\ref{M2}) to write [{\it cf.} Eq. (\ref{C3})]
\begin{equation}
\delta\mu_{j}=\frac{e}{i\omega C_{j}}\sum_{\ell}I_{\ell j}.\label{C6}
\end{equation}
We then obtain
\begin{eqnarray}
I_{\ell_{1}\ell_{2}}&=&\frac{z_{\ell_{1}\ell_{2}}}{e}
\Bigl ( i\omega\varphi_{\ell_{1}\ell_{2}}\nonumber\\
&+&\frac{e}{i\omega C_{\ell_{1}}}\sum_{j}I_{j\ell_{1}}
-\frac{e}{i\omega C_{\ell_{2}}}\sum_{j}I_{\ell_{1}j}\Bigr ).\label{C7}
\end{eqnarray}
This can be rearranged to yield
\begin{eqnarray} \label{ke}
I_{\ell_{1}\ell_{2}}\Biggl (1&+&\frac{z_{\ell_{1}\ell_{2}}}{i\omega}
\Bigl (\frac{1}{C_{\ell_{1}}}+\frac{1}{C_{\ell_{2}}}\Bigr )\Biggr )+
\frac{z_{\ell_{1}\ell_{2}}}{i\omega C_{\ell_{2}}}
\sum_{\stackrel{j}{j \neq \ell_{1}}}I_{j\ell_{2}}\nonumber\\
&+&\frac{z_{\ell_{1}\ell_{2}}}{i\omega C_{\ell_{1}}}
\sum_{\stackrel{j}{j \neq \ell_{2}}}I_{\ell_{1}j}
=\frac{i\omega}{e}\varphi_{\ell_{1}\ell_{2}}z_{\ell_{1}\ell_{2}}.\label{C8}
\end{eqnarray}
Equation (\ref{C8}) represents an array of ${\cal L}\times{\cal L}$
equations for the bond currents, where ${\cal L}$ is the number of bonds in
the system. We shall not attempt to present the formal general
solution for this array.  
Instead, we will obtain the limiting behaviors for the frequency dependence
of $\vec{M}$, and exemplify them by considering the smallest possible
cluster of 
sites, a triangle, which possesses a magnetic moment. (For a two-site
cluster the 
orbital magnetization (\ref{M7}) vanishes because
$[\vec{r}_{1}\times\vec{r}_{2}]=0$ 
by construction.)

Inspection of Eqs. (\ref{C8}) reveals the appearance of the 
characteristic frequency $\omega_{0}^{j\ell}$, given by
\begin{equation}
\omega_{0}^{j\ell}=\frac{z_{j\ell}}{C_{j(\ell )}}.\label{C9}
\end{equation}
When the external frequency $\omega$ is much higher than all 
$\omega_{0}^{j\ell}$'s, the local chemical potentials are ineffective
in determining the bond currents. In that case
\begin{equation}
I_{\ell j}\sim \frac{i\omega}{c}\varphi_{\ell j}z_{\ell j}.\label{C10}
\end{equation} 
On the other hand, when $\omega $ is less than all $\omega_{0}^{j\ell}$'s,
the local chemical potentials dominate the bond currents. In that situation
[cf. Eq. (\ref{C6})] the current in the $j\ell $ bond ``feels" the
fluxes on all other bonds, and the solution of Eqs. (\ref{C8}) is
much more complicate. For example, for a three-site cluster we find
\begin{equation}
I_{12}=I_{23}=I_{31}=\frac{i\omega\varphi}{e}\Bigl (\frac{1}{z_{12}}
+\frac{1}{z_{23}}+\frac{1}{z_{31}}\Bigr )^{-1},\label{C11}
\end{equation}
where $\varphi $ is the total flux enclosed by the triangle
\begin{equation}
\varphi =\varphi_{12}+\varphi_{23}+\varphi_{31}.\label{C12}
\end{equation}
That is, $i\omega\varphi /c$ is the e.m.f. applied to the cluster.
Recalling that for a triangle
$\vec{r}_{1}\times\vec{r}_{2}=\vec{r}_{2}\times\vec{r}_{3}=\vec{r}_{3}
\times\vec{r}_{1}=2\vec{S}/3$,
where $\vec{S}$ is the (vectorial) area of the three-site group, 
we find that the magnetic moment can be written in the form
\begin{equation}
M_{\text{triangle}}=
\frac{i\omega}{c^{2}}HS^{2}z_{j\ell}, \label{C13}
\end{equation}
where $z_{j\ell}$ stands for the {\it largest} admittance of the three in the
high frequency case, and for the {\it smallest} one at very low frequencies.
Note, however, that for a triangle, the numerical coefficient in the
high frequency limit includes a geometrical factor ($\varphi_{12}
/3\varphi$)  
which diminishes the
ratio $M/\omega $.

Let us now focus on the frequency dependence of the admittance itself. 
We consider the
cases where $\omega\ll|\epsilon_{j\ell}| $ and
$\omega\gg|\epsilon_{j\ell}| $. 
One can then ignore the resonant contribution to $\Re z_{j\ell}$.
For $\omega\ll|\epsilon_{j\ell}| $ we have
\begin{equation}
z_{j\ell}=G_{j\ell}(1+i\frac{\omega }{\omega_{1}^{jl}}),\label{C14}
\end{equation}
with $\omega_{1}^{j\ell}$ given by
\begin{equation}
\omega_{1}^{j\ell} \sim \left|\frac{v^{2}{\cal D}\beta\epsilon_{j\ell}}{n_{j}
-n_{\ell}}\right| e^{-\frac{\beta}{2}(|\epsilon_{j}| +|\epsilon_{\ell}|
+|\epsilon_{j\ell}| )}\sim 
e^{-\beta|\epsilon_{j\ell}| }.\label{C15}
\end{equation}
In the other extreme limit, where $\omega\gg|\epsilon_{j\ell}| $, the 
imaginary part of the admittance becomes inversely proportional to
$\omega $. When 
the electron-phonon coupling is weak, $z_{j\ell}$ 
in this regime is dominated by the imaginary part, leading to
\begin{equation}
z_{j\ell}\sim iZ_{j\ell}/\omega, \ \ Z_{j\ell}=e^{2}| J_{j\ell}|^{2}
(n_{\ell}-n_{j})/\epsilon_{\ell j}.\label{C16}
\end{equation}

Collecting the above results, we obtain the following pattern for the
behavior 
of $M_{\text{triangle}}$ as function of the temperature and the external 
frequency. Starting
at very low frequencies, such that $\omega <|\epsilon_{j\ell}|$ and
$\omega <\omega_{1}^{j\ell}$, the admittance is mainly given by the bond dc
conductance, so that
\begin{equation}
M_{\text{triangle}}\sim\frac{i\omega}{c^{2}}HS^{2}G_{j\ell}.\label{C17}
\end{equation}
The magnetic moment decays exponentially as $T\rightarrow 0$. As the
frequency increases such that $\omega >\omega_{1}^{j\ell}$, then
$z_{j\ell}\sim iG_{j\ell}\omega /\omega_{1}^{j\ell}$ and
\begin{equation}
M_{\text{triangle}}\sim \frac{\omega^{2}}{c^{2}}HS^{2}
G_{j\ell}/\omega_{1}^{j\ell}.\label{C18}
\end{equation}
The leading frequency dependence is modified as compared to the very
low $\omega $ 
regime, but the moment is still exponentially decaying with the
decrease of the 
temperature [see Eqs. (\ref{17}) and (\ref{C15})]. 
Finally, in the high frequency limit such that 
$\omega\gg|\epsilon_{j\ell}| $, the moment becomes almost independent
of the temperature and the frequency
\begin{equation} 
M_{\text{triangle}}\sim Z_{j\ell}HS^{2}/c^2.\label{C19}
\end{equation}
For frequencies in the intermediate  regime the behavior is rather
complicate, and depends upon the detailed relationships between the
frequency, temperature, $|\epsilon_{j\ell}|$, and the characteristic frequencies
$\omega_{0}^{j\ell}$ and $\omega_{1}^{j\ell}$.

\section{Concluding remarks}

The orbital magnetization of localized electrons in response to the
temporal variation of a vector potential has been studied. We have
obtained an expression for the magnetic moment in terms of the
tunneling and phonon-assisted hopping currents between localized states.
These currents have been derived from the Kubo formula.

The mere existence of an orbital magnetic moment in the strongly localized
system requires a group of at least three states. We have found that the
magnetization of this small cluster is proportional to the relevant
dominant admittance between sites (the smallest one for small frequencies, 
$\omega\ll\omega_{0}^{j\ell}$, and the largest one in the opposite case). 
Consequently, $M_{\text{triangle}}$ is proportional to $|J|^{2}$, where $J$
is the overlap integral of two wave functions. The {\it equilibrium}
magnetic moment of a triangle \cite{we}, on the other hand, is of order
$J^{3}$. This is because $M_{\text{eq}}$ results from quantum
interference processes 
among the three sites.

The magnetization of a bulk system involves summation over all bonds [see 
Eq. (\ref{M7})]. Its full calculation requires the knowledge of the 
distribution function of the site energies and the localized site locations.
This may lead to a temperature dependence different from the exponential one
characterizing the small cluster. Consider for example the case of
nearest-neighbor hopping \cite{e-sh} 
at low enough frequencies. Then the typical energies 
involved are within a region of width $\sim \ \beta^{-1}$ around the
Fermi level. 
Consequently, for a smooth distribution of the site energies, within
an energy 
band of width $W$, the number of triangles contributing to the
magnetic moment 
is $\propto (\beta W)^{-3}$. The typical spatial separation between sites is
$\sim \ \xi L_{\beta}$, with $\xi $ being the localization length and
$L_{\beta}\propto \ln |\beta J_{0}|$. ($J_{0}$ is the bare overlap
integral between  
two localized levels.) It follows that the area of a typical triangle is of 
order $(\xi L_{\beta})^{2}$. Collecting these estimates and using
Eq. (\ref{C17}) 
we obtain that the bulk magnetic moment {\it per unit volume} is
\begin{equation}
M_{\text{bulk}}\sim i\omega (\frac{e}{c})^{2}Hv^{2}{\cal D}(\beta^{-1})\beta
(\frac{n_{i}}{\beta W})^{3}(\xi L_{\beta})^{8}\xi ^{2}. \label{D1}
\end{equation}
Here $n_{i}$ is the concentration of localized levels.
The bulk moment is proportional to the frequency and depends on the
temperature $T$ as $T^2{\cal D}(T)\ln^8(J_0/T)$. 
Modifications, due e.g.
to the temperature dependence of the relevant area in the case of
variable-range hopping, are possible. Finally we remark that in this work
electronic correlations were ignored \cite{e-sh}. 
These have to be taken into account
in the calculation of the bond currents. We hope to pursue this issue in
a future article.

\acknowledgements
We thank Y. Imry and I. Goldhirsh for helpful discussions. Support 
by the German-Israel Fund (GIF),
and the Fund for Basic Research administered by the Israel 
Academy of Sciences and Humanities
is acknowledged. We gratefully acknowledge partial support 
from the joint Israel-Norway Program for Cultural Exchange.

\appendix

\section{Quantum interference effects in the current}

As is mentioned above, quantum interference between the direct
amplitude $j\rightarrow \ell$ 
and the indirect one, via a third site $k$, appears first in order
$J^{3}$. It is therefore 
necessary to derive the corrections to the current to third order in
the overlaps. We denote 
this quantity by $I_{jl}^{(3)}$. Returning to Eq. (\ref{7}) we find
\wide
\begin{eqnarray}
I_{j\ell}^{(3)}(t)&=&-e\int_{-\infty}^{t}dt_{1}\int_{-\infty}^{t_{1}}dt_{2}
\left<\Biggl[\Bigl[{\cal H}_{1}(t_{2}),{\cal H}_{\text
{per}}(t_{1})\Bigr]\,
2\Im J_{\ell j}Q_{\ell j}(t)c_{\ell}^{\dagger}(t)c_{j}(t)\Biggr]\right>
\nonumber\\ & & \quad \quad
-\int_{-\infty}^{t}dt_{1}\int_{-\infty}^{t}dt_{2}
\left<\Biggl[{\cal H}_{\text {per}}(t_{1}), \Bigl[
{\cal H}_{1}(t_{2}),
2\Im J_{\ell j}Q_{\ell
j}(t)c_{\ell}^{\dagger}(t)c_{j}(t)\Bigr]\Biggr]\right>
\nonumber\\&&
 -\phi_{\ell
j}(t)\Biggl\{\int_{-\infty}^{t}dt_{1}\int_{-\infty}^{t_{1}}dt_{2} 
\left<{\cal H}_{1}
(t_{2}){\cal H}_{1}(t_{1})
2\Re J_{\ell j}Q_{\ell j}(t)
c_{\ell}^{\dagger}(t)c_{j}(t)
+2\Re J_{\ell j}Q_{\ell j}(t)c_{\ell}^{\dagger}(t)c_{j}(t){\cal H}_{1}(t_{1})
{\cal H}_{1}(t_{2})\right>
\nonumber\\ && \quad \quad
+\int_{-\infty}^{t}dt_{1}\int_{-\infty}^{t}dt_{2}\left<{\cal H}_{1}(t_{1})
2\Re J_{\ell j}Q_{\ell j}(t)c_{\ell}^{\dagger}(t)c_{j}(t){\cal
H}_{1}(t_{2})\right>\Biggr\}.\eqnum{A1}\label{A1} 
\end{eqnarray}
The calculation of the terms appearing in Eq. (\ref{A1}) is 
straightforward, though rather cumbersome. 
It yields for the ac amplitude of $I_{j\ell}^{(3)}$, 
$\bar{I}_{j\ell}^{(3)}$, the expression
\begin{eqnarray}
\bar{I}_{j\ell}^{(3)}(\omega)&=&
2e\sum_{k}\int_{0}^{\infty}d\tau_{1}\int_{0}^{\infty}d\tau_{2}\Biggl\{
F_{k\ell,kj}(-\tau_{1},\tau_{2})
\Bigl[\bar{\phi}_{\ell j}
+\bar{\phi}_{k\ell}
e^{-i\omega\tau_{1}}+\bar{\phi}_{jk}e^{-i\omega\tau_{2}}\Bigr]n_{k}
\bar{n}_{j}\bar{n}_{\ell}
-F_{jk,\ell k}(-\tau_{1},\tau_{2})
\nonumber \\ && \times 
\Bigl[\bar{\phi}_{\ell j}
+\bar{\phi}_{jk}
e^{-i\omega\tau_{1}}
+\bar{\phi}_{k\ell}e^{-i\omega\tau_{2}}\Bigr]\bar{n}_{k}n_{j}n_{\ell}
-F_{j\ell,jk}(-\tau_{1},-\tau_{2})
\Bigl[\bar{\phi}_{\ell j}
+\bar{\phi}_{k\ell}
e^{-i\omega\tau_{1}}+\bar{\phi}_{jk}e^{-i\omega(\tau_{1}+\tau_{2})}\Bigr]
n_{j}\bar{n}_{k}\bar{n}_{\ell}
\nonumber\\&&
+F_{j\ell, k\ell}(-\tau_{1},-\tau_{2})
\Bigl[\bar{\phi}_{\ell j}
+\bar{\phi}_{jk}
e^{-i\omega\tau_{1}}
+\bar{\phi}_{k\ell}e^{-i\omega(\tau_{1}+\tau_{2})}\Bigr]n_{j}n_{k}
\bar{n}_{\ell}
-F_{\ell j,\ell k}(\tau_{1},\tau_{2})
\Bigl[\bar{\phi}_{\ell j}
+\bar{\phi}_{jk}
e^{-i\omega\tau_{1}}
\nonumber \\ &&
+\bar{\phi}_{k\ell}e^{-i\omega(\tau_{1}+\tau_{2})}
\Bigr]
n_{\ell}\bar{n}_{j}\bar{n}_{k}
+F_{\ell j,kj}(\tau_{1},\tau_{2})\Bigl[\bar{\phi}_{\ell j}
+\bar{\phi}_{k\ell}e^{-i\omega\tau_{1}}
+\bar{\phi}_{jk}e^{-i\omega(\tau_{1}+\tau_{2})}\Bigr]n_{\ell}n_{k}
\bar{n}_{j}\Biggr\},
\eqnum{A2}\label{A2}
\end{eqnarray}
where $\bar{n}_{j}=1-n_{j}$, and
\begin{equation}
F_{\ell j, kj}(\tau_{1},\tau_{2})=\Re\Bigl\{J_{jk}J_{k\ell}J_{\ell j}
e^{i\epsilon_{\ell j}\tau_{1}+i\epsilon_{kj}\tau_{2}
+g(\tau_{1}+\tau_{2})+g(\tau_{1})+g(\tau_{2})-3g(0)}\Bigr\}.
\eqnum{A3} \label{A3}
\end{equation}
Contrary to the behavior of $\bar{I}_{j\ell}^{(2)}(\omega )$,
$\bar{I}_{j\ell}^{(3)}(\omega  
=0)$ is finite. Setting $\omega =0$ in Eq. (\ref{A2}) one finds that
each $k-$term in the sum is proportional to 
\begin{equation}
\bar{\phi}_{\ell jk}=\bar{\phi}_{\ell
j}+\bar{\phi}_{jk}+\bar{\phi}_{k\ell},
\eqnum{A4} \label{A4} 
\end{equation}
that is, to the total flux enclosed in the triangle $\ell jk$. The
zero frequency 
part of the current is just the persistent current 
flowing in response to a constant magnetic flux. Its properties in the
strong localization 
regime have been studied in great detail elsewhere \cite{we}.
Denoting 
\begin{equation}
\bar{I}_{j\ell}^{(3)}(\omega =0)=I_{pc}, 
\eqnum{A5} \label{A5}
\end{equation}
we derive from (\ref{A2}) $\bar{I}_{j\ell}^{(3)}(\omega )-I_{pc}$ for
weak electron-phonon 
interaction, and in the small $\omega $ limit. The result is
\begin{eqnarray}
\bar{I}_{j\ell}^{(3)}(\omega )-I_{pc}&=&i\omega
e\sum_{k}J_{jk}J_{k\ell}J_{\ell j}
\Biggl\{\bar{\phi}_{k\ell}\Biggl[\Bigl(\frac{1}{\epsilon_{\ell
j}}+\frac{1}{\epsilon_{kj}}\Bigr)\beta 
n_{k}\bar{n}_{\ell}\gamma_{k\ell}+\Bigl(\frac{1}{\epsilon_{jk}}+
\frac{1}{\epsilon_{\ell k}}\Bigr)\beta n_{\ell}\bar{n}_{j}\gamma_{\ell
j}\Biggr]
+\bar{\phi}_{jk}\Biggl[\Bigl(\frac{1}{\epsilon_{k\ell}}+\frac{1}
{\epsilon_{j\ell}}\Bigr)\beta n_{j} 
\bar{n}_{k}\gamma_{jk}
\nonumber\\&+&
\Bigl(\frac{1}{\epsilon_{jk}}+\frac{1}{\epsilon_{\ell k}}\Bigr)\beta
n_{\ell}\bar{n}_{j}\gamma_{\ell j}\Biggr]
+2i\omega\Biggl[\bar{\phi}_{k\ell}\frac{1}{\epsilon_{kj}}
\Bigl(\frac{n_{k}-n_{\ell}}
{\epsilon_{\ell k}^{3}}+\frac{n_{\ell}-n_{j}}{\epsilon_{\ell
j}^{3}}\Bigr)
+\bar{\phi}_{jk}\frac{1}{\epsilon_{\ell
k}}\Bigl(\frac{n_{k}-n_{j}}{\epsilon_{kj}^{3}}+ 
\frac{n_{j}-n_{\ell}}{\epsilon_{\ell j}^{3}}\Bigr)\Biggr]\Biggr\}.
\eqnum{A6} \label{A6}
\end{eqnarray}
\narrow
A remarkable observation is that $\bar{I}_{l\ell}^{(3)}(\omega )$,
which is part of the 
current in the $j\ell $ bond, is driven by the magnetic phases on the bonds
$k\ell $ and $jk$, but {\it not} by the phase on the bond $j\ell $
itself. This 
is a manifestation of interference effects. The interference affects 
other electronic properties of the system. One example is the
conductivity. The formalism 
presented here can be used to find that quantity. In that case the electric
field $\vec{E}$ is 
related to the fluxes by
\begin{equation}
i\omega\bar{\phi}_{j\ell}\rightarrow e\vec{E}\cdot\vec{R}_{j\ell}.
\eqnum{A7} \label{A7}
\end{equation}
Then the ring currents $I_{pc}$ vanish by virtue of (\ref{A4}). 
Using Eq. (\ref{A7}) one finds that
Eqs. (\ref{13}), and (\ref{A6}) give the conductivity of
the $j\ell $-bond.  
In particular, Eqs. (\ref{13}) and (\ref{17}) yield the well-known
result for the bond conductivity 
to second order in the overlaps, while Eq. (\ref{A6}) gives the {\it
nonlocal} corrections 
to it, which result from interference \cite{holstein,we2}.
Namely, the current in the $j\ell $-bond is
affected by the voltage drops on the $k\ell $ and the $jk$ bonds.

\widetext
\end{document}